\documentclass[12pt]{article}
\pdfoutput=1

\usepackage[margin=3.5cm]{geometry}
\usepackage{amsmath,amssymb,graphicx}
\usepackage[dvipsnames]{xcolor}
\usepackage{hyperref}
\usepackage{multicol}
\usepackage{soul}
\usepackage{slashed}
\usepackage{appendix}
\usepackage{changepage}
\usepackage{comment}
\usepackage{authblk}
\usepackage{scalefnt}

\renewcommand{\geq}{\geqslant}

\numberwithin{equation}{section}

\begin{document}

%%%%%%%%%%%%%%%%%%%%%%%%%%%%%%%%%%%%%%%%%%%%%%%%%%%%%%%%%%%%

\title{Photon Scattering by an Electric Field in Noncommutative Spacetime}

\author{Daniela D'Ascanio, Pablo Pisani, Ulises Wainstein Haimovichi}

\affil{Instituto de Física La Plata (IFLP; UNLP-CONICET)\\ La Plata, Argentina}

\date{\today}

\maketitle

\thispagestyle{empty}

\begin{abstract}
	As is known, the existence of a small noncommutativity between coordinates would generate nonlocal self-interactions in the electromagnetic theory. To explore some consequences of this effect on the propagation of photons we consider Moyal space half-filled with a static and homogeneous electric field and analyze electromagnetic fluctuations on top of this step-like background. Both the localization of photons and the possibility of photon production by strong electric fields are addressed. Several aspects of the Klein paradox in this setup are discussed as well.
\end{abstract}

\vfill

\noindent\rule{5cm}{.25mm}

\noindent{\footnotesize dascanio@fisica.unlp.edu.ar\\
pisani@fisica.unlp.edu.ar\\
ulises2357@hotmail.com}

\pagebreak

\tableofcontents

\vspace{7mm}
\hrule
\vspace{2mm}

\section{Introduction}

As a rule, photons do not interact among themselves. Nevertheless, such self-interaction is possible in noncommutative (NC) spacetimes. Moyal spacetime is a most natural NC scenario where noncommutativity between coordinates can be represented by the Moyal $\star$-product between fields \cite{Douglas:2001ba,Szabo:2001kg}. Electrodynamics in this spacetime, known as $U(1)_\star$, has been extensively studied: as a first remarkable feature one finds that noncommutativity introduces nonlinearities in Maxwell's equations through a self-interaction which is nonlocal already at the classical level. In other words, since gauge fields---as functions of NC coordinates---do not commute, $U(1)_\star$ holds a close resemblance with nonabelian Yang-Mills theories (see the recent lecture notes \cite{Vitale:2023znb}).

The goal of the present work is to expose some aspects of this self-interaction in the scattering of photons by a static electric field background. One of our original motivations was the question of whether noncommutativity allows photon creation by intense electromagnetic fields (the effects of noncommutativity on electron/positron production by strong electric fields can be found in \cite{Chair:2000vb}). Although the answer requires a quantum field theoretic approach, the full set of stationary solutions contained in this article lays down the ingredients for a subsequent analysis of Schwinger effect in this framework (either through the S-matrix formalism \cite{Nikishov:2001ps,Gavrilov:2015yha} or with heat-kernel techniques \cite{Vassilevich:2003xt}).

In spite of the similarity between $U(1)_\star$ and commutative Yang-Mills theories, constant chromoelectric fields do create gluons \cite{Nayak:2005yv} whereas homogeneous fields do not produce photons in NC space (some aspects of photon propagation on homogeneous electromagnetic fields have been analyzed in \cite{Ilderton:2010rx} and \cite{Fresneda:2015zya}). For this reason, we study the propagation of photons across the interface of two regions, one with and the other without a background electric field. For simplicity, we take an electrostatic potential that only varies along a fixed spatial direction, and such that an homogeneous electric field is suddenly built up: NC space gets then split into halves by a flat interface. Next, we study the propagation of fluctuations of this background.

More specifically, our setting is the following. We start with $(d+1)$-dimensional Moyal spacetime, whose coordinates satisfy $[\hat x^\mu,\hat x^\nu]=2i\theta^{\mu\nu}$, where $\theta^{\mu\nu}$ are elements of a real antisymmetric matrix. For simplicity, we take the time coordinate $x^0$ as an ordinary commuting parameter, i.e., $\theta^{\mu0}=0$. We then introduce a particular background, namely, a static electric field which vanishes on the left half-space $\hat x^1<0$, takes a constant value on the right half-space $ \hat x^1>0$, and is perpendicular to the interface $\hat x^1=0$. The potential $A_0=-E(|\hat x^1|+\hat x^1)/2$ generates such a step-like field; the constant $E$ gives the electric field.

Due to the self-interaction introduced by noncommutativity, photons are scattered by this background. However, as will be shown, in the Moyal representation the interface effectively acts as a step of finite width with (depending on the polarization) delta-functions at the boundaries. This step scatters oblique incident photons; perpendicular beams do not experience any interaction with the electric field. According to the intensity of the background field, the energy and the direction of the incident photons one recognizes different regimes. Apart from the usual scattering---where incident photons split into partially transmitted and reflected light beams---one also finds, for a background which is not particularly strong, the propagation of edge states (viz.\ localized along the interface). Bound states as perturbations of solitonic solutions have already been found in \cite{Vassilevich:2003he}.

On the other hand, for extremely intense electric fields (roughly, $|E|\sim 1/|\theta^{\mu\nu}|$) a realization of the Klein paradox arises. Our result for the scattering of vector particles is similar to the usual (i.e., commutative) Klein paradox for scalar particles \cite{Winter}: the reflection coefficient $R$ for a wave packet is greater than one. This would indicate that at the boundary of regions with strong electric fields photon beams could be produced.

There is an interesting result of our calculations that concerns two different manifestations of the Klein paradox. As is well-known, a reflection coefficient $R$ greater than one is found either for monochromatic solutions of Dirac equation or for localized solutions of Klein-Gordon equation. In both cases this is considered as a sign of particle creation. At the same time, one finds (both for spinor and scalar particles) the paradoxical result that the transmission coefficient $T$ does not vanish for infinitely strong barriers \cite{Calogeracos:1998rf}. In the problem studied in this article we find $R>1$ (for a localized packet) but, on the other hand, $T\to 0$ as the background field grows to infinity. This shows that these two aspects are not necessarily connected. It also suggests the need to perform a second quantized approach to reliably determine the rate of photon production.

Another aspect which is usually considered in the Klein regime is the divergence of the transmission coefficient for some value of the incident energy; this is usually called super-Klein tunneling (see e.g.\ \cite{Kim}). For a bosonic field colliding against an abrupt step this occurs at incident energy equal to half the step height. In our setting there is no super-Klein tunneling, i.e., the transmission coefficient remains finite. The reason lies in the smearing introduced by noncommutativity which turns the abrupt edge into a region of finite slope (cf.\ \cite{Winter}).

All these features of photon propagation are analyzed in the present article both for TE as well as for TM modes; sometimes called s-polarized and p-polarized waves, respectively. The former are not affected by gauge transformations; for the latter we choose the temporal gauge. Interestingly, this gauge choice introduces a singular background in the equation of motion. We show that this is a general picture in the Klein regime---independently of the background profile. However, the origin of the singularity can be clearly identified and its consequences removed.

Since its discovery \cite{K}, the Klein paradox has been thoroughly analyzed from different perspectives. Many aspects of its relation with particle creation in quantum field theory have been thoroughly discussed---we mention, e.g., \cite{Nikishov:1970br,Hansen:1980nc,Manogue,Dombey:1999id,Chervyakov:2011nr,Evans:2021coy}. More recently, techniques based on the space-time resolved solutions have provided further insight into the behavior of such systems (see, e.g., \cite{KSG,CSG,Lv:2018wpn,Alkhateeb:2022yjn}). We think that the example considered in this article gives interesting information on the Klein paradox in a different setup, namely, one with propagating photons under nonlocal interactions. Moreover, for a specific photon polarization, we are lead to the problem of a single scalar field colliding against a step of finite width\footnote{This is the scalar version of Sauter's seminal paper \cite{Sauter}.}. We consequently analyze some aspects of the solutions which, to the best of our knowledge, have not been studied before (the explicit solutions were already written in \cite{Gavrilov:2015zem}).

Our article is organized as follows. In Section \ref{Moyal} we briefly review the construction of $U(1)_\star$ and focus on the dynamics of fluctuations around a classical background. In Section \ref{tb} we introduce our specific background and perform the decoupling between TE and TM modes. In Section \ref{regimes} we identify the different regimes according to the energy of the incident beam. Sections \ref{tc} and \ref{ss} are devoted to the study of TE and TM modes, respectively. Singularities arising in the Klein region are resolved in Section \ref{kz}. In Section \ref{es} we analyze the existence of edge states. Finally, in Section \ref{conclu}, we draw our conclusions.

\section{Electrodynamics in Moyal spacetime}\label{Moyal}

A natural procedure to set up a $(d+1)$-dimensional NC spacetime is to introduce operators $\hat x^\mu$ (with $\mu=0,...,d$) such that
\begin{align}
	[\hat x^\mu,\hat x^\nu]=2i\theta^{\mu\nu}\,.
\end{align}
The real constants $\theta^{\mu\nu}$ represent minimal measurable areas, according to the uncertainty relation $\Delta x^\mu\Delta x^\nu\geq |\theta^{\mu\nu}|$. We adopt the mostly minus signature $\eta=(+-\ldots-)$.

Coordinate operators $\hat x^\mu$ generate a NC algebra of classical fields $\phi(\hat x)$ that can be represented in the space of ordinary functions $\phi(x)$ on $\mathbb{R}^{d+1}$ through the Weyl-Wigner transform,
\begin{align}
\phi(\hat x)=\int \frac{d^{d+1}x\,d^{d+1}p}{(2\pi)^{d+1}}\ e^{ip(\hat x-x)}\ \phi(x)\,.
\end{align}
This establishes a one-to-one map between operators $\phi(\hat x)$ and functions $\phi(x)$. We use $\hat x^\mu$ to denote NC coordinates and unhatted $x^\mu$ for the ordinary commuting coordinates of Moyal spacetime. One can check that under this correspondence the composition of NC fields is mapped into the Groenewold-Moyal product of functions,
\begin{align}\label{mp}
	\phi(\hat x)\cdot \psi(\hat x)\leftrightarrow
	\phi(x)\star \psi(x)
	%=\phi(x)\ e^{i\theta^{\mu\nu}\overleftarrow\partial_\mu\overrightarrow\partial_\nu}\ \psi(x)
	&=\phi(x^\mu+i\theta^{\mu\nu}\partial_\nu)\, \psi(x)\nonumber\\[2mm]
	&=\psi(x^\mu-i\theta^{\mu\nu}\partial_\nu)\, \phi(x)\,.
\end{align}
This representation can be readily checked, for example, on the original commutator $[x^\mu,x^\nu]=x^\mu\star x^\nu-x^\nu\star x^\mu=2i\theta^{\mu\nu}$.

It can also be shown that any cyclic trace in the algebra of operators $\phi(\hat x)$ gives, up to a normalization factor, the integral on $\mathbb{R}^{d+1}$ of the corresponding functions $\phi(x)$ \cite{Szabo:2001kg}. Therefore, any action functional $S[\phi(\hat x)]$ can be written as the integral of a Lagrangian density obtained by replacing every composition of operators with the $\star$-product of ordinary commuting fields. As a consequence, the NC generalization of local field theories turns out to be nonlocal because interactions become, roughly speaking, smeared over distances of order $\sqrt{|\theta^{\mu\nu}|}$. As we will see, in the case of electrodynamics invariance under nonlocal gauge transformations leads to a self-interacting theory.

As in the commutative case, the nonlocal symmetry $U(1)_\star$ is implemented through the covariant derivative
\begin{align}
	D_\mu=\partial_\mu+iA_\mu
\end{align}
that introduces a gauge field $A_\mu(x)$. Throughout this article we will work in the adjoint representation so the action of $D_\mu$ should be understood as $[D_\mu,\cdot\,]$, where, of course, the commutator is defined with the $\star$-product. The commutator $[D_\mu,D_\nu]=i[F_{\mu\nu},\cdot]$ defines the covariant field strength
\begin{align}
	F_{\mu\nu}=\partial_\mu A_\nu-\partial_\nu A_\mu+i\,[A_\mu,A_\nu]\,,
\end{align}
which clearly shows the resemblance between electromagnetism in Moyal spacetime and ordinary Yang-Mills theories.

As briefly explained above, the generalization of Maxwell action to Moyal spacetime is simply given by
\begin{align}
	S[A]=\int d^{d+1}x
	\ \left(-\frac14\,F^{\mu\nu}\star F_{\mu\nu}-J^\mu\star A_\mu\right)\,.
\end{align}
Upon integration by parts the $\star$-product can be removed from quadratic terms in the action but it must be retained in cubic and quartic self-interactions. To introduce in our setting a background electric field we have included a fixed external source $J^\mu(x)$.

The equation of motion
\begin{align}\label{source}
	D_\mu F^{\mu\nu}=J^\nu
\end{align}
determines a configuration of the gauge field for each external source. Still, an immediate implication of \eqref{source} is that the current must be covariantly constant,
\begin{align}\label{divJ}
	D_\mu J^\mu=0\,.
\end{align}

Let us now assume a specific solution $A_\mu(x)$ and introduce small fluctuations $a_\mu(x)$ around this background. If we write the full gauge field as $A_\mu(x)+a_\mu(x)$, the (linearized) field tensor becomes
\begin{align}\label{changeFmunu}
	F_{\mu\nu}+D_\mu a_\nu-D_\nu a_\mu\,.
\end{align}
In this expression and hereafter, $F_{\mu\nu}$ and $D_\mu$ are computed exclusively in terms of the background field $A_\mu$. Replacing \eqref{changeFmunu} into \eqref{source} we obtain the linearized equation of motion for the perturbations,
\begin{align}\label{eom-nogauge}
	-\eta_{\mu\nu}\,D^2 a^\nu+D_\mu D_\nu a^\nu-2i\,[F_{\mu\nu},a^\nu]=0\,.
\end{align}
Note also that, since $J^\mu$ is the same for $A_\mu$ and $A_\mu+a_\mu$, condition \eqref{divJ} now implies
\begin{align}\label{aconjota}
	[a_\mu,J^\mu]=0\,.
\end{align}

Before considering in detail a specific background, let us examine gauge invariance on the perturbations. In the presence of a background field, gauge invariance remains as a small $U(1)_\star$ symmetry on the fluctuations,
\begin{align}\label{gauge}
	\delta a_\mu=D_\mu\alpha(x)\,,
\end{align}
that preserves their equation of motion as long as
\begin{align}\label{condition}
	0=[D_\mu F^{\mu\nu},\alpha]=[J^\nu,\alpha]\,.
\end{align}
Therefore, we can conveniently gauge the perturbations with a parameter $\alpha(x)$ that commutes with the external source.

\section{The background}\label{tb}

Let us define our specific setup. As already mentioned, we are interested on the effects of spatial noncommutativity, so we take the time coordinate $x^0$ as a commuting variable. As for the space coordinates, note that after an appropriate choice the antisymmetric matrix $\theta^{\mu\nu}$ can be diagonalized in $2\times2$ blocks. In this way, space becomes decomposed into several Moyal planes such that coordinates from different planes commute with each other. For our purposes it is enough to distinguish the coordinates $x^1,x^2$ of one of these planes from the remaining $d-2$ spatial coordinates. Noncommutativity is then characterized by a positive parameter $\theta$ through
\begin{align}
	[x^1,x^2]=2i\theta
\end{align}
together with a $(d-2)\times (d-2)$ antisymmetric matrix in the rest of Moyal space. As will become evident soon, if the background does not depend on these $d-2$ coordinates then this antisymmetric  matrix is irrelevant.

We now introduce a background $A_0=A_0(x^1)$ that varies only along the direction of $x^1$, and set the spatial components to zero: $A_1=\ldots=A_d=0$. We will shortly choose a specific $x^1$-dependence for $A_0$. This static background represents a purely electric field $E(x^1)=-A_0'(x^1)$ in the $x^1$--direction, and originates on a static distribution of charge density $J^0=-A_0''(x^1)$; spatial components of the current vanish as well, $J^1=\ldots=J^d=0$.

Since the static background does not change with the transverse coordinates $x_\perp=(x^2,\ldots,x^{d})$, we analyze fluctuations that propagate with fixed energy $k^0$ and transverse momentum $k_\perp=(k^2,\ldots,k^{d})$,
\begin{align}\label{waves}
	a_\mu(x)=a_\mu(x^1)\,e^{-ik^0x^0+ik_\perp x_\perp}\,.
\end{align}
Here, $k_\perp x_\perp=k^2x^2+\ldots+k^dx^d$.

It is now important to point out that the commutator between any background which only depends on $x^1$ and any field with definite momentum $k^2$ can be written as an ordinary multiplication between the same field and another function which, roughly, evaluates the variation of the original function at distances $\Delta x^1\sim \theta k^2$. For example,
\begin{align}\label{adjoint}
	[A_0(x^1),a_\mu(x^1)\,e^{-ik^0x^0+ik_\perp x_\perp}]=
	\delta\! A_0(x^1)\,a_\mu(x^1)\,e^{-ik^0x^0+ik_\perp x_\perp}\,,
\end{align}
with
\begin{align}\label{delta}
	\delta\! A_0(x^1)=A_0(x^1-\theta k^2)-A_0(x^1+\theta k^2)\,.
\end{align}
This expression shows that if $k_\perp$ has no component in the direction of $x^2$ the background is transparent to photon beams. As already mentioned, since the background only depends on $x^1$, the propagation of photons will only be affected by $\theta$, and the remaining NC parameters will have no physical consequence.

\begin{figure}
	\centering
	\begin{minipage}{.8\linewidth}
		\includegraphics[width=10cm]{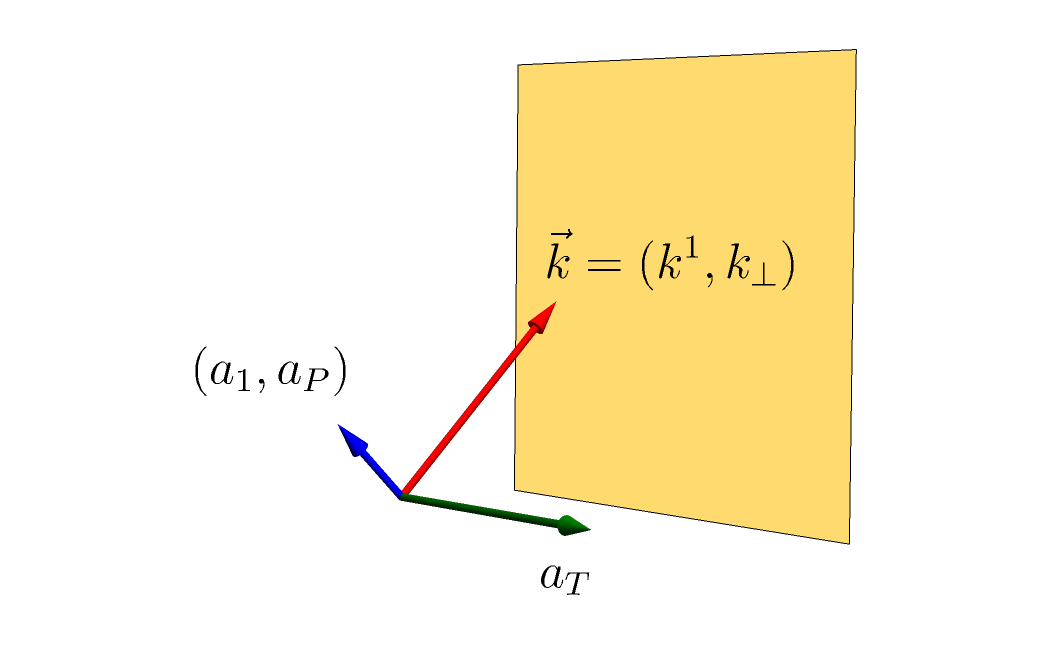}
		\caption{\small Decomposition of the gauge field into TE modes $a_T$ (green) and TM modes $a_1,a_P$ (blue).}
		\label{flechas}
	\end{minipage}\hspace{10mm}
	
\end{figure}

Before solving \eqref{eom-nogauge} for our specific background, let us establish some convenient notation. We decompose the fluctuations $a_\mu=(a_0,\ldots,a_d)=(a_0,a_1;a_\perp)$ as follows. In an asymptotic region---where the background eventually vanishes---the fluctuations propagate with momentum $\vec k=(k^1,...,k^d)=(k^1;k_\perp)$; as expected, in the absence of a background the wave equation \eqref{eom-nogauge} indicates that the spatial part of $a_\mu$ is perpendicular to $\vec k$. We separate from the transverse field $a_\perp$ two components: (i) the projection $a_P$ in the plane of incidence (i.e., the component parallel to $k_\perp$), and (ii) $a_T$, which is a $(d-3)$-component vector perpendicular to the direction of $x^1$ and to $k_\perp$.  This decomposition is illustrated in figure \ref{flechas} for an incoming wave entering a region of non-vanishing background (represented by the flat surface).
With some abuse of notation,
\begin{align}
	a_\mu=a_0\oplus a_1\oplus a_P \oplus a_T\,.
\end{align}
In the asymptotic region the electric field of an incoming wave is proportional to the spatial components of the gauge field. Therefore, $a_1,a_P$ describe a beam with electric field in the plane of incidence (TM mode or p-polarized wave). On the other hand, the components $a_T$ give an electric field perpendicular to the plane of incidence (TE mode or s-polarized wave).

Through the equation of motion \eqref{eom-nogauge} each component of $a_T$ gets completely decoupled,
\begin{align}
	\left[(k_0-\delta\! A_0)^2-|k_\perp|^2+\partial_1^2\right] a_T=0\,.\label{muT}
\end{align}
TM modes then satisfy Klein-Gordon equation for a charged field of mass $|k_\perp|$ in interaction with an electrostatic potential $\delta\!A_0(x^1)$. After choosing a specific background $\delta\!A_0$ we will solve this equation in section \ref{tc}.

The remaining components $a_0(x^1),a_1(x^1),a_P(x^1)$ satisfy the coupled equations
\begin{align}
	a_0''-|k_\perp|^2a_0
	-(k_0-\delta\! A_0)\left(ia_1'-|k_\perp|\,a_P\right)+2i\delta\! A_0'\,a_1&=0\,,\label{mu0}\\[2mm]
	%%%
	\left[(k_0-\delta\! A_0)^2-|k_\perp|^2\right] a_1
	+ i(k_0-\delta\! A_0)a_0'-i|k_\perp|a_P'+i\delta\! A_0'\,a_0&=0\,,\label{mu1}\\[2mm]
	%%%
	(k_0-\delta\! A_0)^2a_P+a_P''
	-|k_\perp| \left[(k_0-\delta\! A_0) a_0+ia_1'\right]&=0\,.\label{mu2}
\end{align}
These equations imply condition \eqref{aconjota}, which can now be written as $\delta\!J^0\, a_0=0$ or, equivalently, as $\delta\!A_0''\, a_0=0$. Under this condition any two equations among \eqref{mu0}-\eqref{mu2} imply the third one. Consequently, we will choose the temporal gauge $a_0=0$, which trivially satisfies \eqref{aconjota}, and will use \eqref{mu0} and \eqref{mu1} to find $a_1$ and $a_P$. This will be done in section \ref{ss}.

\begin{figure}
	\begin{minipage}[c]{.38\textwidth}
		\centering
		\includegraphics[width=6cm]{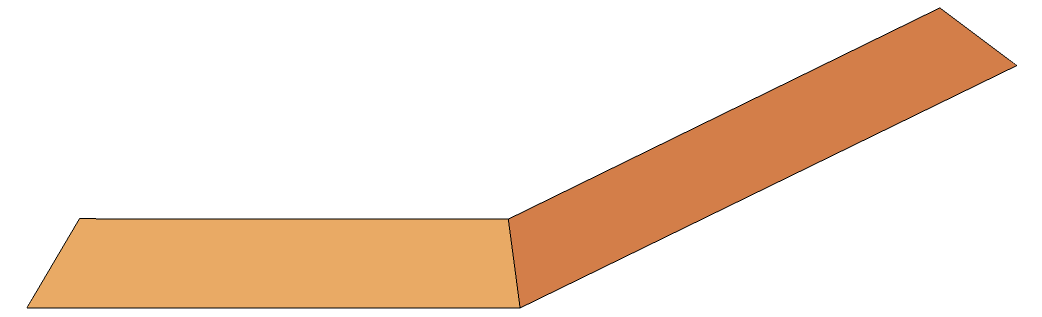}
		\caption{\small Background slope from the NC spacetime perspective. In the picture $E<0$ so the electric field points towards the left.}
		\label{step}
	\end{minipage}
	\hspace{1.25cm}
	\begin{minipage}[c]{.515\textwidth}
		\centering
		\includegraphics[width=7.5cm]{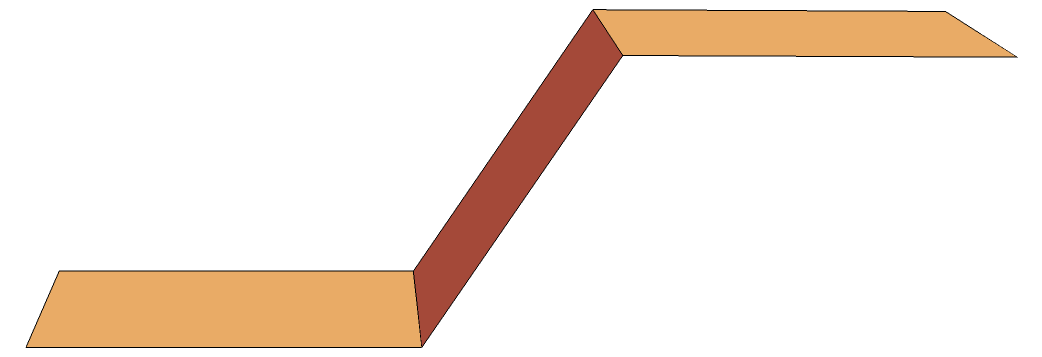}
		\caption{\small The background as it interacts with incoming photons (for ${\rm sign}(k^2)E>0$). The width of the step depends on the transverse momentum $k^2$; its slope, on the electric field $E$.}
		\label{well}
	\end{minipage}
\end{figure}

Now it is time to choose a specific profile for the background $A_0(x^1)$: We take a slope that begins abruptly at $x^1=0$ (as in figure \ref{step}),
\begin{align}\label{back}
	A_0(x^1)=\left\{
	\begin{array}{llll}
		0&&&x^1<0\\[2mm]
		-E\,x^1&&&x^1>0
	\end{array}
	\right.\,.
\end{align}
The parameter $E\in\mathbb{R}$ gives the intensity of the electric field. From the viewpoint of the NC spacetime, this potential represents the boundary of a region supporting a static homogeneous electric field. This setting can be considered as a first approach to the study of the dynamics of fluctuations at the interface where an electric field is built up.

Recall that, according to \eqref{adjoint}, the action of $A_0(x^1)$ on oscillations \eqref{waves} turns the slope-like potential into a 
broadened step of small width $2L$ (figure \ref{well}), with
\begin{align}
	L=\theta |k^2|\,.
\end{align}
In fact, inserting \eqref{back} into \eqref{delta} one obtains
\begin{align}\label{a0}
	\delta\! A_0(x^1)
	=\left\{
	\begin{array}{lll}
		0&&x^1<-L\\[2mm]
		-E(x^1+L)&&|x^1|<L\\[2mm]
		-2EL&&x^1>L
	\end{array}
	\right.\,.
\end{align}
In this expression and in the rest of this article we assume ${\rm sign}(k^2)<0$; if the component $k^2$ is positive one simply changes the sign of $E$. Correspondingly, the adjoint action of the electric field is given by
\begin{align}\label{a0p}
	[F_{01},\,\cdot\,]=-\delta\! A_0'(x^1)
	=\left\{
	\begin{array}{lll}
		0&&|x^1|>L\\[2mm]
		E&&|x^1|<L
	\end{array}
	\right.\,.
\end{align}
This shows that photons effectively interact with the electric field only in a narrow region of width $2L$.

Correspondingly, the homogeneous background on the half-plane is generated by a charge density $J^0=\frac12E\,\delta(x^1)$ together with a second parallel plate of opposite charge density and located at $x^1=+\infty$. Condition $\delta J^0\,a_0=0$ now forces $a_0$ to vanish at the edges of the step.

\section{Energy regimes}\label{regimes}

Fluctuations \eqref{waves} are characterized by their energy $k_0$ and their transverse momentum $k_\perp$. As already mentioned, photons incident normally to the interface---or even with no $k^2$ component---do not interact with the background. Let us call $\beta$ the angle between the plane of incidence and the $x^1x^2$-plane. One can distinguish, both for TE and TM modes, different energy regimes according to the intensity of the electric field: (i) very strong electric fields such that $\theta|E|\cos{\beta}>1$, and (ii) ordinary electric fields for which $\theta|E|\cos{\beta}<1$.

\begin{figure}
	\centering
	\begin{minipage}{.45\linewidth}
	\includegraphics[width=7cm]{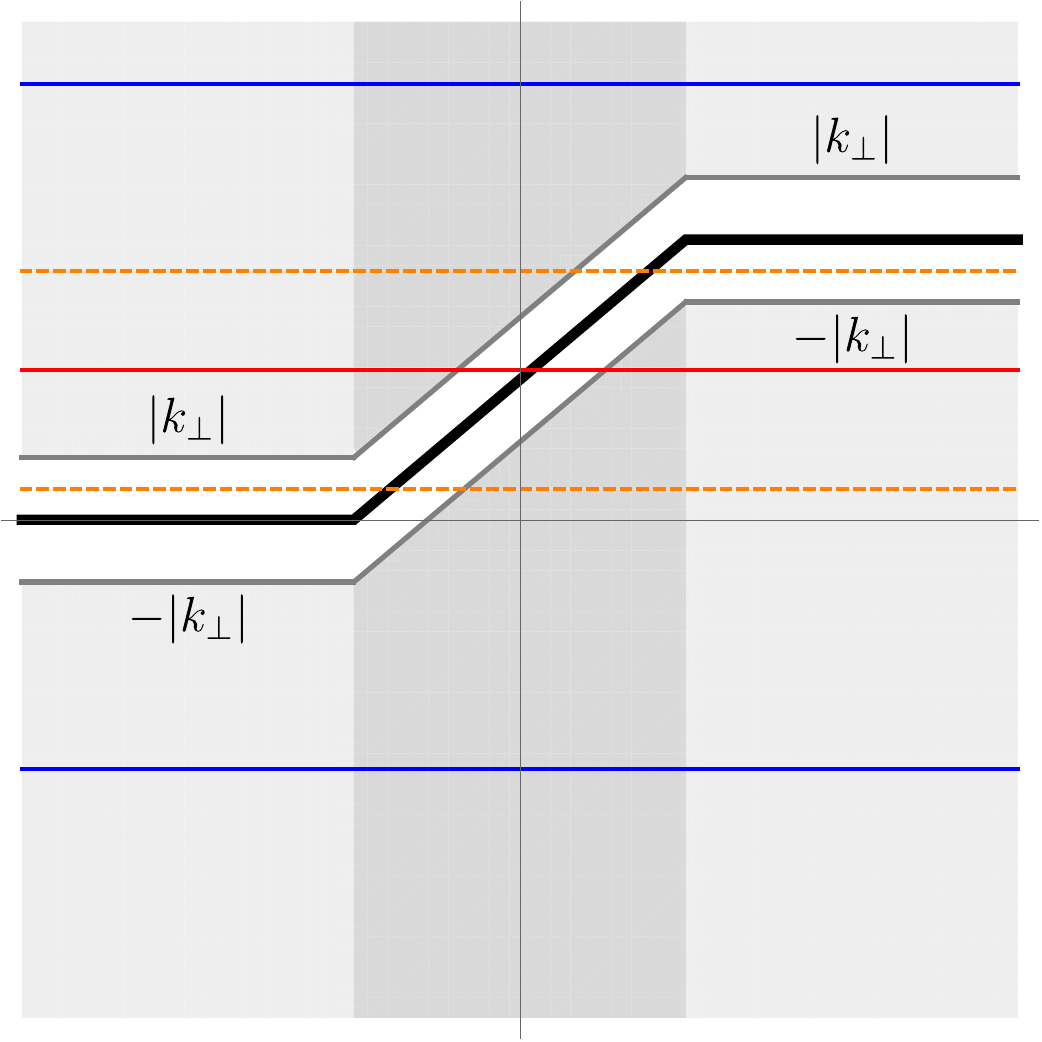}
	\caption{\small The black solid line represents the potential $\delta\!A_0$ for a strong electric field $E<0$. Horizontal lines correspond to five different energy regimes.}
	\label{zones2}
	\end{minipage}\hspace{10mm}
	\begin{minipage}{.45\linewidth}
	\includegraphics[width=7cm]{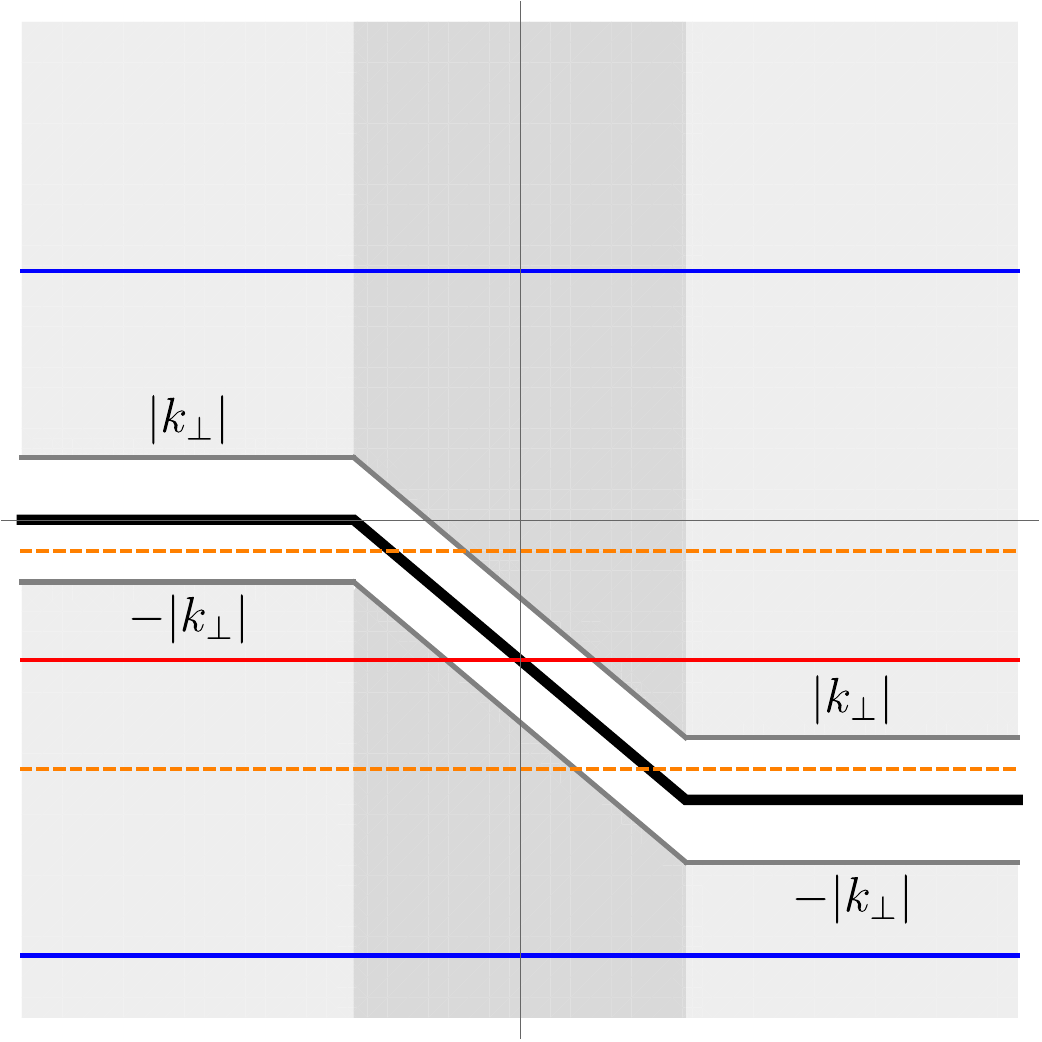}
	\caption{\small The black solid line represents the potential $\delta\!A_0$ for a strong electric field $E>0$. Horizontal lines correspond to five different energy regimes.}
	\label{zones1}
	\end{minipage}
\end{figure}

Case (i) is represented in figures \ref{zones2} and \ref{zones1} for $E<0$ and $E>0$, respectively. The transverse momentum $k_\perp$ introduces a mass gap. Accordingly, five different energy regimes can be identified: the blue lines represent photons with positive or negative energy that collide against the step-like background and generate reflected and transmitted waves; this we call the {\it propagation region}. Orange dashed lines represent photons which propagate only at one side of the interface with an energy that lies within the mass gap at the other side. Therefore, the photon does propagate across the interface and suffers a total reflection. Finally, there exists a finite interval of energies, usually called {\it Klein zone}, for which the incident photon lies above and below the surface of the Dirac sea, depending on the side of the step. This is represented by the red horizontal lines in the figures. In this regime particles (with positive kinetic energy) at one side of the step are seen as antiparticles (with negative kinetic energy) at the other side.

In case (ii) one also finds the usual propagation region but the electric field is not strong enough to separate the mass gaps and generate a Klein zone. As a consequence, there is an overlapping between the mass gaps at either side of the interface so that one could eventually find states localized close to this surface (edge states). This possibility will be analyzed in section \ref{es}.

\section{TE modes}\label{tc}

We begin by solving equation of motion for TE modes $a_T$. Recall that each of the $d-2$ components---which we now generically denote as $\phi(x^1)$---satisfies the Klein-Gordon equation for a charged scalar field of mass $|k_\perp|$ scattered by an electrostatic potential $\delta\!A_0$. Taking into account the piecewise definition of $\delta\!A_0(x^1)$, we solve \eqref{muT} by appropriately matching at $x^1=\pm L$ the solutions in the left region $x^1<-L$, the right region $x^1>L$, and the intermediate region $|x^1|<L$. These we denote by $\phi_L$, $\phi_R$ and $\phi_I$, respectively.

In the regions $|x^1|>L$ the background $\delta\!A_0$ is constant so the solutions are simply incoming and outgoing plane waves at either side of the step,
\begin{align}
	\phi_L(x^1)&=a\, e^{ik^1x^1}+b\, e^{-ik^1x^1}\qquad {\rm for\ }x^1<-L\,,\label{phiL}\\[2mm]
	\phi_R(x^1)&=c\, e^{i\kappa^1 x^1}+d\, e^{-i\kappa^1 x^1}\qquad {\rm for\ }x^1>L\,,\label{phiR}
\end{align}
with
\begin{align}
	k^1&=+\sqrt{k_0^2-|k_\perp|^2}\,,\label{kauno}\\[2mm]
	\kappa^1&=+\sqrt{\left(k_0+2EL\right)^2-|k_\perp|^2}\,.\label{kappa}
\end{align}
In the intermediate region $|x^1|<L$ the function $\delta\!A_0$ is linear in $x^1$ so the field satisfies
\begin{align}
	-\phi_I''+\left(|k_\perp|^2-E^2z^2\right)\phi_I=0\,,\label{tmequ}
\end{align}
where we have shifted the $x^1$ coordinate as
\begin{align}\label{z}
	z=x^1+L+\frac{k_0}{E}\,.
\end{align}
The solutions can be written as
\begin{align}
	\phi_I(x^1)=A\, F(x^1)+B\, G(x^1)\qquad {\rm for\ }-L<x^1<L\,,\label{phiI}
\end{align}
in terms of parabolic functions
\begin{align}
	F(x^1)=\frac{\Gamma(-\nu)}{\sqrt{2i|E|}}&\left[D_\nu(-\sqrt{2i|E|}\,\tfrac{k0}{E})D_\nu(\sqrt{2i|E|}z)
	-\mbox{}\right.\nonumber\label{FTE}\\[2mm]
	&\left.\mbox{}-D_\nu(\sqrt{2i|E|}\,\tfrac{k0}{E})D_{\nu}(-\sqrt{2i|E|}z)\right]\,,\\[2mm]
	G(x^1)=\Gamma(-\nu)&\left[D'_\nu(-\sqrt{2i|E|}\,\tfrac{k0}{E})D_\nu(\sqrt{2i|E|}z)
	+\mbox{}\right.\nonumber\\[2mm]
	&\left.\mbox{}+D'_\nu(\sqrt{2i|E|}\,\tfrac{k0}{E})D_{\nu}(-\sqrt{2i|E|}z)\right]\,,\label{GTE}
\end{align}
where $\nu=-1/2+ik_\perp^2/2|E|$. These particular combinations where chosen to satisfy $F(-L)=G'(-L)=0$ and $F'(-L)=G(-L)$.

After imposing continuity of $\phi$ and its derivative at $x^1=\pm L$ one obtains four equations: two of them give the coefficients $A,B$ in the intermediate region in terms of $a,b,c,d$; the remaining two equations determine the elements of the S-matrix,
\begin{align}\label{sm}
	S=\left(\begin{array}{cc}
		r_L	&	t_R	 \\ t_L	&	r_R
	\end{array}\right)\,,
\end{align}
that relates the outgoing coefficients $b,c$ with the incoming coefficients $a,d$ through
\begin{align}
	\left(\begin{array}{c}
		\sqrt{k^1}\,b \\ \sqrt{\kappa^1}\,c
	\end{array}\right)=
	S
	\left(\begin{array}{c}
		\sqrt{k^1}\,a	\\	\sqrt{\kappa^1}\,d	
	\end{array}\right)\,.\label{s}
\end{align}
A straightforward calculation gives
\begin{align}
	r_{L,R}&=\frac{i\kappa^1\left(ik^1F_L\pm G_L\right)\mp ik^1F'_L-G_L'}
	{i\kappa^1\left(ik^1F_L-G_L\right)-ik^1F'_L+G_L'}\ e^{-2ik_{L,R}L}\,,
	\label{rs}\\[2mm]
	t_{L,R}&=\frac{\sqrt{8\pi k^1\kappa^1}\,i}
	{i\kappa^1\left(ik^1F_L-G_L\right)-ik^1F'_L+G_L'}\ e^{-i(k_L+k_R) L}\,.\label{ts}
\end{align}
The upper (lower) sign corresponds to $r_L$ ($r_R$). We have also defined $k_L=k^1$ and $k_R=\kappa^1$. We use $F_L,G_L$ to denote $F(L),G(L)$, and similarly for the derivatives.

Relation \eqref{s} parametrizes the two-dimensional space of solutions. The coefficients $a$ and $d$ represent incoming waves from the left and from the right, respectively. Thus, if $d=0$ then $b$ is related to the reflection coefficient and $c$ to the transmission coefficient of a wave colliding the step from the left. Correspondingly, $a=0$ describes a wave colliding from the right and the roles of $b$ and $c$ get interchanged. To define reflection and transmission coefficients we compute the conserved current\footnote{This current corresponds to the conserved (canonical) momentum of the electromagnetic field. This issue is discussed in section \ref{conclu}.}
\begin{align}\label{jota}
	J(x^1)={\rm Im}\,(\phi^*\phi')\,,
\end{align}
which according to \eqref{muT} must be $x^1$-independent. For $x^1<-L$, expression \eqref{phiL} gives
\begin{align}
	J&={\rm Re}(k^1)
	\Big[|a|^2 e^{-2x^1{\rm Im}(k^1)}
	-|b|^2 e^{2x^1{\rm Im}(k^1)}\Big]
	-2\,{\rm Im}(k^1)\,{\rm Im}\Big[ab^*e^{2ix^1{\rm Re}(k^1)}\Big]\,.
\end{align}
The same expression holds for $x^1>L$ if one makes the following replacements: $k^1\to\kappa^1$, $a\to c$ and $b\to d$.

For energies in the propagation and in the Klein regimes both $k^1$ and $\kappa^1$ are positive real numbers, so current conservation $J(-\infty)=J(+\infty)$ implies
\begin{align}\label{cc}
	k^1|a|^2-k^1|b|^2=\kappa^1 |c|^2-\kappa^1 |d|^2\,.
\end{align}
This expression---valid for arbitrary values of the constants $a,d$---proves that $S$ is a unitary matrix. Further relations between the elements of the S-matrix can be obtained by noting that the complex conjugates of \eqref{phiL} and \eqref{phiR} provide another stationary solution but where the roles of $b,c$ and $a,d$ are played by $a^*,d^*$ and $b^*,c^*$, respectively. Therefore, $S^*=S^{-1}=S^\dagger$. This implies $t_R=t_L$ and $|r_L|=|r_R|$. Although these relations can be readily checked in \eqref{rs} and \eqref{ts}, their derivation did not make use of the specific form of the background. In fact, this is a well-known result from one-dimensional scattering: reflection (and transmission) coefficients for left- or right-incident waves coincide.

\begin{figure}[t]
	\centering
	\begin{minipage}{.45\linewidth}
		\includegraphics[width=6.5cm]{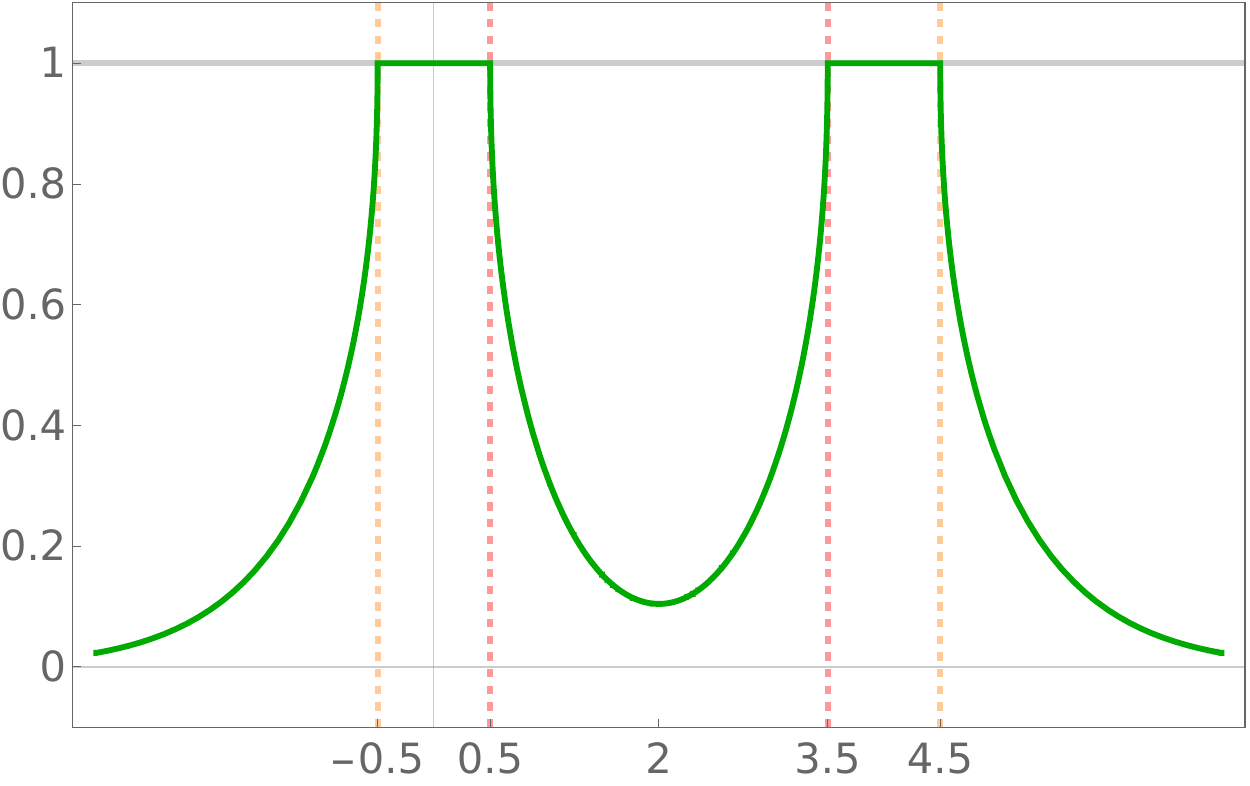}
		\caption{\small $R$ vs.\ $k_0$ for monochromatic TE modes ($|E|\theta\cos{\beta}>1$ and $E<0$). In this picture $k_2=0.25$, $k_\perp=0.5$, $E=-8$ in length units of $\sqrt\theta$.}
		\label{TE_RyT}
	\end{minipage}\hspace{10mm}
	\begin{minipage}{.45\linewidth}
		\includegraphics[width=6.5cm]{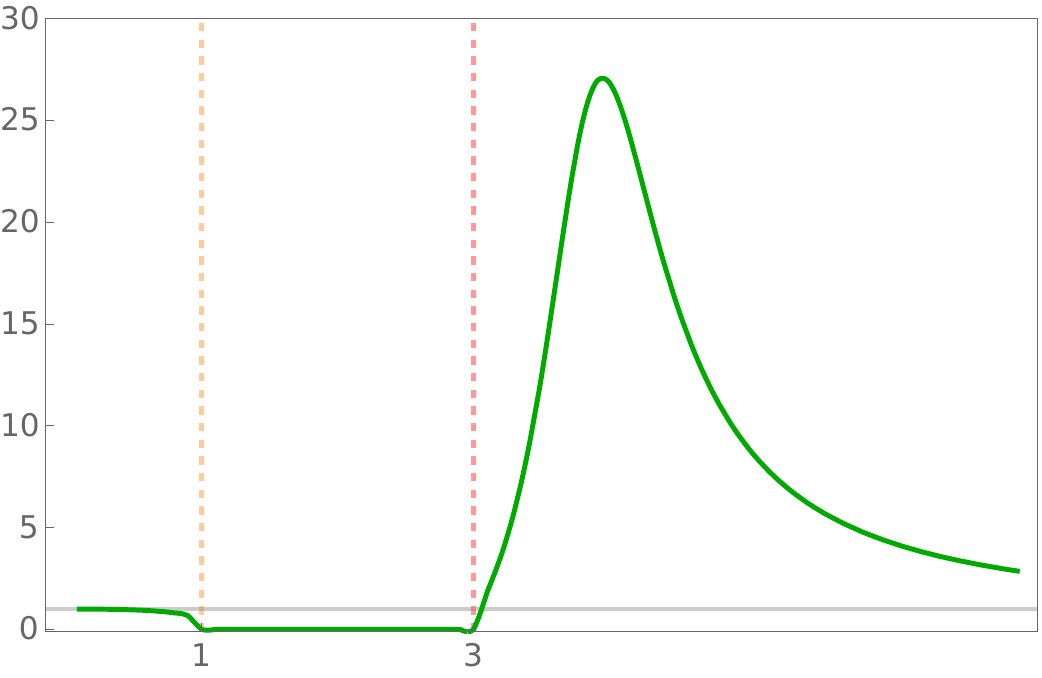}
		\caption{\small $T$ vs.\ $|E|$ for a pulse of TE modes. At the left of the orange dashed line lies the propagation zone; at the right of the red dashed line, the Klein zone. In this picture $E<0$, $k_2=0.25$, $k_\perp=0.5$, $k_0=1$ in length units of $\sqrt\theta$.}
		\label{KleinKleinTE}
	\end{minipage}
\end{figure}

Reflection and transmission coefficients are then defined as $R=|r_L|^2=|r_R|^2$ and $T=|t_L|^2=|t_T|^2$. Unitarity of $S$ implies $R+T=1$. Figure \ref{TE_RyT} displays $R$ as a function of the incident energy $k_0$.
The reflection coefficient decreases monotonically from 1 to 0 with increasing $|k_0|$. If the energy lies within one of the mass gaps then either $k^1$ or $\kappa^1$ are pure imaginary so $J=0$ and the incoming photon is totally reflected at the interface ($R=1$).

In the Klein zone $R$ is smaller than 1 but to get an appropriate interpretation of the reflection coefficient it is crucial to recall that for these energies phase and group velocities have opposite signs. Let us consider, for example, the case $E<0$ (figure \ref{zones2}). For incident energies $|k_\perp|<k_0<-2EL-|k_\perp|$ both momenta $k^1$ and $\kappa^1$ are real but the group velocity at $x^1>L$ is
\begin{align}
	v=\frac{dk_0}{d\kappa^1}=\frac{\kappa^1}{k_0+2EL}<0\,.
\end{align}
Therefore, the stationary solutions which are relevant to model the scattering of a left-incident wave packet are not obtained by choosing $d=0$ in \eqref{phiL} and \eqref{phiR} but instead by taking $c=0$. The reflection and transmission coefficients are then given by $R=1/|r_L|^{2}$ and $T=|t_L|^2/|r_L|^2$ and the relation $R=1+T>1$ holds. This is the Klein paradox which is interpreted as the result of a total reflection of the incident packet together with the creation of pairs of photons at the interface at a rate $T$.

Alternatively, the term ``Klein paradox" is sometimes used to refer to the existence of a non-vanishing transmission coefficient even for a infinitely high barrier---this is the case for a Klein-Gordon or Dirac particle colliding an ordinary abrupt step. Interestingly, this does not happen in our setting. In fact, the leading behavior of \eqref{FTE} and \eqref{GTE} for large $|E|$ reads \cite{A-S}
\begin{align}
	F(x^1)&\sim -\Gamma(\tfrac14)\,z^{-\frac12}\,|E|^{-\frac34}
	\,\sin{(\tfrac12 z^2|E|+\tfrac\pi8)}\,,\\[2mm]
	G(x^1)&\sim -2\Gamma(\tfrac34)\,z^{-\frac12}\,|E|^{-\frac14}
	\,\cos{(\tfrac12 z^2|E|-\tfrac\pi8)}\,.
\end{align}
Consequently, the behavior of $T$ for large values of the electric field is given by
\begin{align}\label{teinfinito}
	T&=\frac{|t_L|^2}{|r_L|^2}
	\sim\frac{2\pi k^1}
	{\left[\Gamma(\tfrac34)\right]^2\sqrt{|E|}}\,.
\end{align}
The dependence of $T$ on the electric field $E$ is shown in figure \ref{KleinKleinTE} for a wave packet. In the propagation region $0<T<1$. If the energy of the packet falls in the mass gap then $T=0$. For strong backgrounds, $T$ may be even greater than 1---the transmitted pulse is interpreted as photons created at the step. Nevertheless, $T\to 0$ as $|E|\to\infty$.

Finally, let us recall that for an ordinary scalar field pulse colliding against an abrupt step (i.e., a step of zero width) the coefficients $R$ and $T$ diverge if the energy is half the step height---this is usually known as super-Klein tunneling. However, the reflection coefficient shown in figure \ref{TE_RyT} (for a monochromatic wave) does never vanishes, so the corresponding coefficient for the wave packet (obtained by replacing $R\to 1/R$) does not diverge. This is a consequence of the smearing of the step in Moyal space. In the limit case $\theta\to0$ and $|E|\to\infty$ (with $|E|\theta$ finite) one obtains and abrupt step and the reflection coefficient for a monochromatic wave certainly vanishes if its energy is exactly half the step height and one recovers super-Klein tunneling in this specific limit.

\section{TM modes}\label{ss}

Components $a_T$ are unaffected by a gauge transformation \eqref{gauge} but to determine $a_0,a_1,a_P$ we need to fix a specific gauge. We solve \eqref{mu0}-\eqref{mu2}  by setting $a_0(x^1)=0$ with an appropriate parameter of the form $\alpha(x)=\alpha(x^1)\,e^{-ik^0x^0+ik_\perp x_\perp}$. Condition \eqref{condition} requires $\alpha(x^1)$ to vanish at the edges of the step $x^1=\pm L$ but, since any solution $a_0(x^1)$ vanishes at those points, one can consistently choose $\alpha$ such that $a_0+D_0\alpha=0$.

For $a_0=0$ equations \eqref{mu0} and \eqref{mu1} read
\begin{align}
	-(k_0-\delta\! A_0)\left(i a_1'-|k_\perp|\,a_P\right)+2i\,\delta\! A_0'\,a_1&=0\,,\label{mu0tg}\\
	\left[(k_0-\delta\! A_0)^2-|k_\perp|^2\right] a_1
	-i|k_\perp|a_P'&=0\label{mu1tg}\,.
\end{align}
As already mentioned, \eqref{mu2} can be derived from these two equations. From \eqref{mu0tg} we eliminate the component which is parallel to the transverse momentum $k_\perp$,
\begin{align}\label{a2ya1}
	-i|k_\perp| a_P
	=a_1'-2\frac{\delta\! A_0'}{k_0-\delta\! A_0}\,a_1\,.
\end{align}
The other component can be written as
\begin{align}\label{a1yphi}
	a_1=\frac{\varphi(x^1)}{k_0-\delta\!A_0}\,,
\end{align}
where $\varphi(x^1)$ satisfies the Schr\"odinger-like equation
\begin{align}\label{kg}
	-\varphi''
	+\left[k_\perp^2
	-(k_0-\delta\!A_0)^2
	+2\frac{\delta\! A_0'\mbox{}^2}{(k_0-\delta\! A_0)^2}
	+\frac{\delta\! A_0''}{k_0-\delta\! A_0}\right]\varphi
	=0\,.
\end{align}

Up to now we have not used the explicit dependence of the background on $x^1$. For the step-like potential \eqref{a0} the fourth term inside the square brackets only introduces delta-functions at the edges of the step,
\begin{align}\label{plasma}
	\delta\! A_0''=E\left[\delta(x^1-L)-\delta(x^1+L)\right].
\end{align}

We now solve \eqref{kg} in the same fashion as for the TE modes: we find a piecewise solution and match the coefficients by imposing the appropriate behavior at the edges of the step. At $x^1<-L$ and $x^1>L$ the field $\varphi$ (respectively denoted by $\varphi_L$ and $\varphi_R$) takes the form
\begin{align}
	\varphi_L=a\, e^{ik^1x^1}+b\, e^{-ik^1x^1}\,,\label{planeL}\\[2mm]
	\varphi_R=c\, e^{i\kappa^1 x^1}+d\, e^{-i\kappa^1 x^1}\,,\label{planeR}
\end{align}
with $k^1$ and $\kappa^1$ once more given by \eqref{kauno} and \eqref{kappa}. At the interval $|x^1|<L$ the field (denoted by $\varphi_I$) satisfies
\begin{align}
	-\varphi_I''
	+\left(|k_\perp|^2-E^2z^2+\frac{2}{z^2}\right)\varphi_I=0\,,\label{fm}
\end{align}
where the variable $z$ is the same as \eqref{z}. It is interesting to make a comparison with the equation for the TM modes \eqref{tmequ}: the dynamics of TE modes contains an additional, singular term $1/z^2$. The effect of this term will be discussed in some detail.

Solutions to \eqref{fm} can be written as combinations of the following confluent hypergeometric functions,
\begin{align}
	f_1(x^1)&=\frac1z\,e^{-\frac i2|E|z^2}\,\mbox{}_1F_1(-\tfrac14-\tfrac{i|k_\perp|^2}{4|E|},-\tfrac12;i|E|z^2)\,,\label{sol1}\\[2mm]
	f_2(x^1)&=z^2\,e^{-\frac i2|E|z^2}\,\mbox{}_1F_1(\tfrac54-\tfrac{i|k_\perp|^2}{4|E|},\tfrac52;i|E|z^2)\,.\label{sol2}
\end{align}
The field at the intermediate region is thus determined by two coefficients $A,B$,
\begin{align}\label{fi}
	\varphi_I=A\,\bar F(x^1)+B\,\bar G(x^1)\,,
\end{align}
where now
\begin{align}\label{f}
	\bar F(x^1)&=-\tfrac13\left[f_2(-L)\,f_1(x^1)-f_1(-L)\,f_2(x^1)\right]\,,\\[2mm]
	\bar G(x^1)&=\tfrac13\left[f'_2(-L)\,f_1(x^1)-f'_1(-L)f_2(x^1)\right]\,.
\end{align}
These particular solutions have been chosen such that $\bar F(-L)=\bar G'(-L)=0$ and $\bar F'(-L)=\bar G(-L)=1$.

Next, we match the behavior of $\varphi_L,\varphi_R,\varphi_I$ at the edges of the step to construct the full solution $\varphi$. We demand continuity of the wave function but, in accordance with \eqref{plasma}, we introduce discontinuities on the derivatives. The matching conditions then read
\begin{align}
	\varphi_L(-L)&=\varphi_I(-L)\,,\label{cmL}\\[2mm]
	\varphi'_L(-L)&=\varphi'_I(-L)+\frac{E}{k_0}\,\varphi_I(-L)\,,\label{cdmL}\\[2mm]
	\varphi_R(L)&=\varphi_I(L)\,,\label{cL}\\[2mm]
	\varphi'_R(L)&=\varphi'_I(L)+\frac{E}{k_0+2EL}\,\varphi_I(L)\,.\label{cdL}
\end{align}
From these four equations we compute the elements $\bar r_L,\bar r_R,\bar t_L,\bar t_R$ of the S-matrix which, as in \eqref{s}, relate outgoing with incoming coefficients,
\begin{align}
	\bar r_{L,R}&=-\frac{\left(\frac{E}{k_0}\mp ik^1\right)\bar F'_L-\bar G_L'
		+\left(\frac{E}{k_0+2EL}\mp i\kappa^1\right)
		\left[\left(\frac{E}{k_0}\mp ik^1\right)\,\bar F_L-\bar G_L\right]}
	{\left(\frac{E}{k_0}+ik^1\right)\bar F'_L-\bar G_L'
		+\left(\frac{E}{k_0+2EL}-i\kappa^1\right)
		\left[\left(\frac{E}{k_0}+ik^1\right)\,\bar F_L-\bar G_L\right]}\, e^{-2ik_{L,R}L}\,,
	\label{rl}\\[2mm]
	\bar t_{L,R}&=\frac{2i\sqrt{k^1\kappa^1}\, e^{-i(k_L+k_R) L}}
	{\left(\frac{E}{k_0}+ik^1\right)\bar F'_L-\bar G_L'
		+\left(\frac{E}{k_0+2EL}-i\kappa^1\right)
		\left[\left(\frac{E}{k_0}+ik^1\right)\,\bar F_L-\bar G_L\right]}\,.\label{tl}
\end{align}
As before, the upper (lower) sign corresponds to $r_L$ ($r_R$); we define $k_L=k^1$ and $k_R=\kappa^1$ and we use $\bar F_L,\bar G_L,\bar F'_L,\bar G'_L$ to denote $\bar F(L),\bar G(L),\bar F'(L),\bar G'(L)$.

\begin{figure}
	\centering
	\begin{minipage}{.45\linewidth}
		\includegraphics[width=6.6cm]{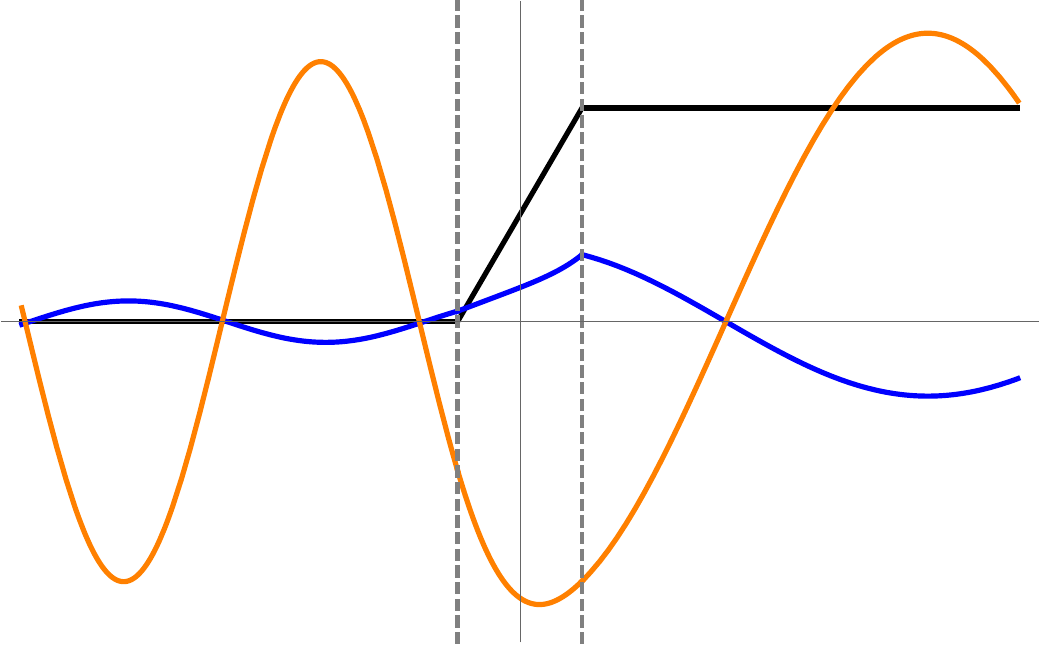}
		\caption{\small Gauge field components $a_1$ (blue) and $a_P$ (orange) with energy in the propagation regime. In black, the background field ($E<0$). The delta-functions at $x^1=\pm L$ introduce discontinuities in the derivatives.}
		\label{pot1}
	\end{minipage}\hspace{10mm}
	\begin{minipage}{.45\linewidth}
		\includegraphics[width=6.6cm]{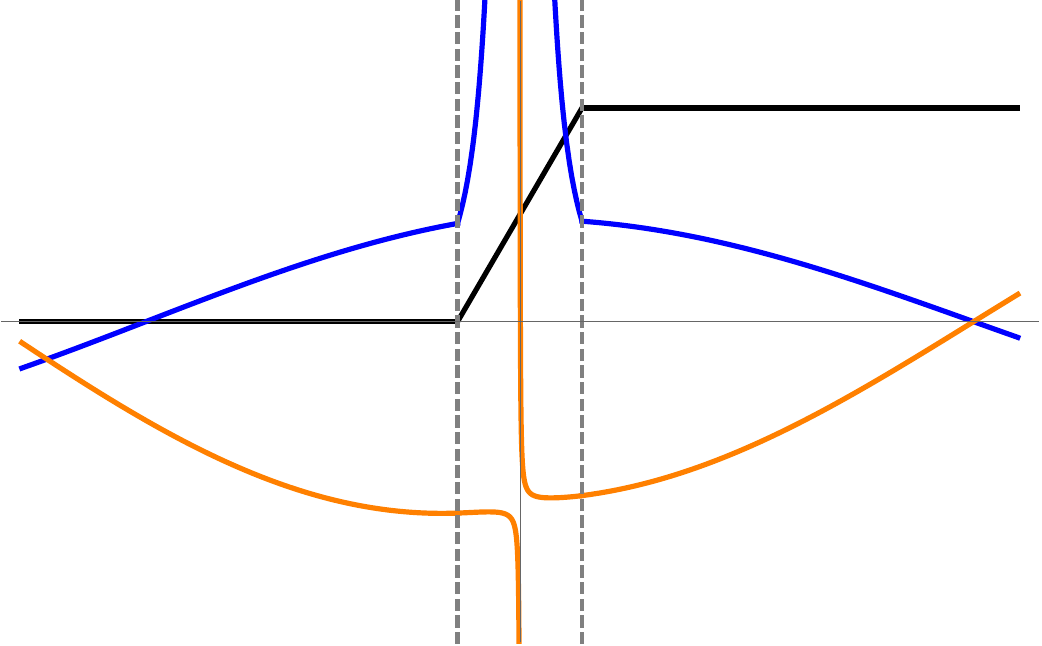}
		\caption{\small Gauge field components $a_1$ (blue) and $a_P$ (orange) with energy in the Klein regime. The background field (black) corresponds to $E<0$. Due to the singularity in \eqref{fm}, the solutions diverge at $z=0$.}
		\label{pot2}
	\end{minipage}
\end{figure}

Figure \ref{pot1} shows the full solution for the gauge field components $a_1(x),a_P(x^1)$ with energy in the propagation regime. As expected, the wave is continuous but its derivative jumps at the edges of the step.
Figure \ref{pot2} displays instead both components $a_1(x^1),a_P(x^1)$ in the case where the energy of the gauge field lies in the Klein zone. The most remarkable feature is the singularity at some $x^1\in[-L,L]$. The obvious reason for this divergence is the singularity in the differential equation \eqref{fm} at $z=0$. For energies in the propagation region $z$ never vanishes in the interval $x^1\in[-L,L]$ but in the Klein regime there is some value of $x^1$ in this interval where $z=0$. In consistence with figure \ref{pot2}, the component $a_1$ has a double pole, whereas $a_P$ has a simple pole. We postpone to section \ref{kz} a more detailed discussion of the physical origin of this divergence and a procedure to appropriately remove it.

\begin{figure}[t]
	\centering
	\includegraphics[width=13cm]{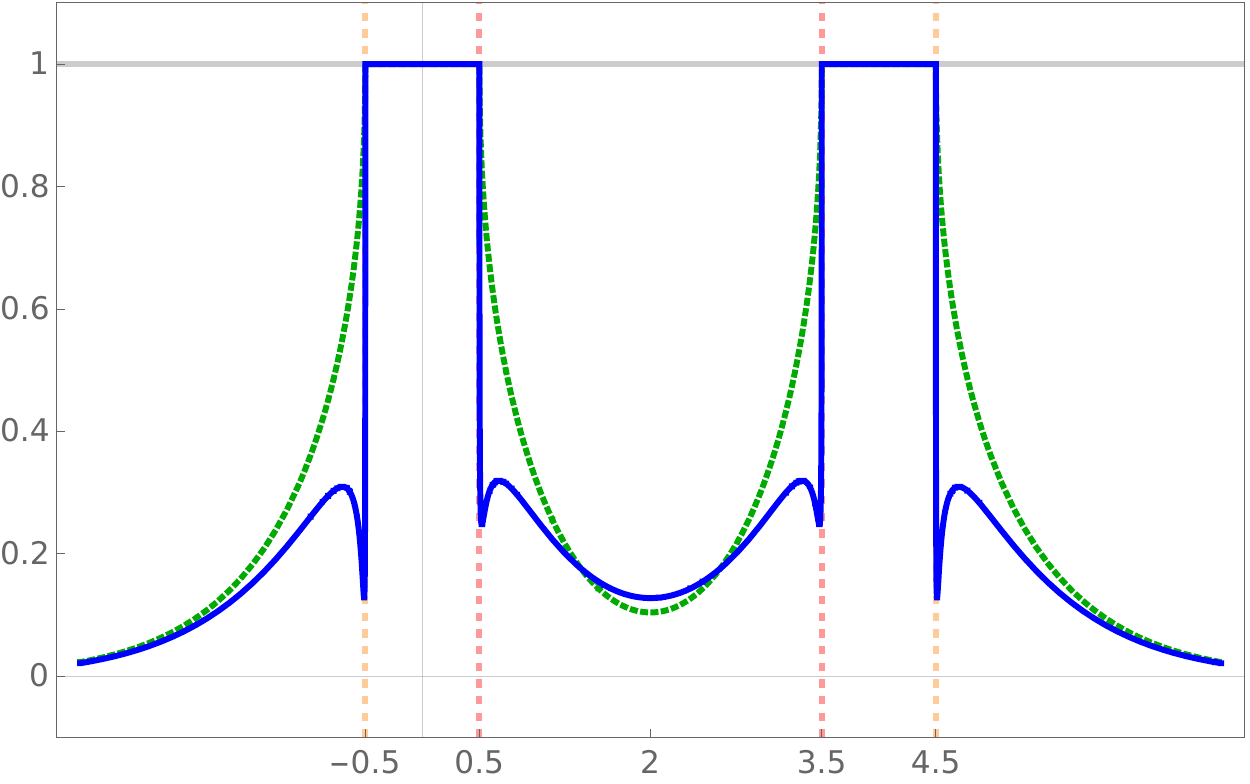}
	\caption{\small $R$ vs.\ $k_0$ (for $|E|\theta>1$ and $E<0$) for a monochromatic TM mode (blue). The interval within the red vertical dashed lines is the Klein zone. The intervals where $R=1$ correspond to the mass gaps. The outer region, delimited by the orange vertical dashed lines is the propagation region. We also display the reflection coefficient for the TE modes (dashed green) as in figure \ref{TE_RyT}. We have chosen $k_2=0.25$, $k_\perp=0.5$, $E=-8$ in length units of $\sqrt\theta$.}
	\label{RyT}
\end{figure}

Let us comment on the solutions \eqref{rl} and \eqref{tl}. We have obtained $t_L=t_R$ and, since $\bar F,\bar G$ are real functions, one readily checks $|r_L|=|r_R|$, as long as $k^1,\kappa^1\in\mathbb{R}$. The reflection coefficient for a monochromatic wave is then given by $R=|r_L|^2$. As before, unitarity of the S-matrix implies $R+T=1$. Figure \ref{RyT} shows the reflection coefficient as a function of the photon energy $k_0$.
The plot is similar to the one corresponding to TE modes but with a sudden decrease in the vicinity of the mass-gap regions, where $R=1$. The figure shows that, in general, TE modes suffer a higher reflection than TM modes.

\section{The Klein zone}\label{kz}

Let us analyze the divergences that occur in the Klein zone for TM modes. The emergence of singularities stems from the covariant derivative $D_0=\partial_0+iA_0(x^1)$ which acting on stationary solutions introduces the factor $k_0-\delta\!A_0(x^1)$. In a typical Klein paradox scenario the potential attains asymptotically constant values $\delta\!A_0(\pm\infty)$; for intense backgrounds their difference $\delta\!A_0(+\infty)-\delta\!A_0(-\infty)$ exceeds the mass gap and thus the band of energies known as Klein zone is created. For $k_0$ within this band there is necessarily at least one point $x^1$ (assuming continuous backgrounds) such that $k_0=\delta\!A_0(x^1)$, so the covariant derivative vanishes at this point. Decoupling the components of the gauge field in \eqref{mu0}-\eqref{mu2} then leads to the appearance of a singular term $(k_0-\delta\!A_0)^{-2}$ in the equation of motion (third term in \eqref{kg}).

\begin{figure}[t]
	\centering
	\begin{minipage}{.4\linewidth}
		\includegraphics[width=6cm]{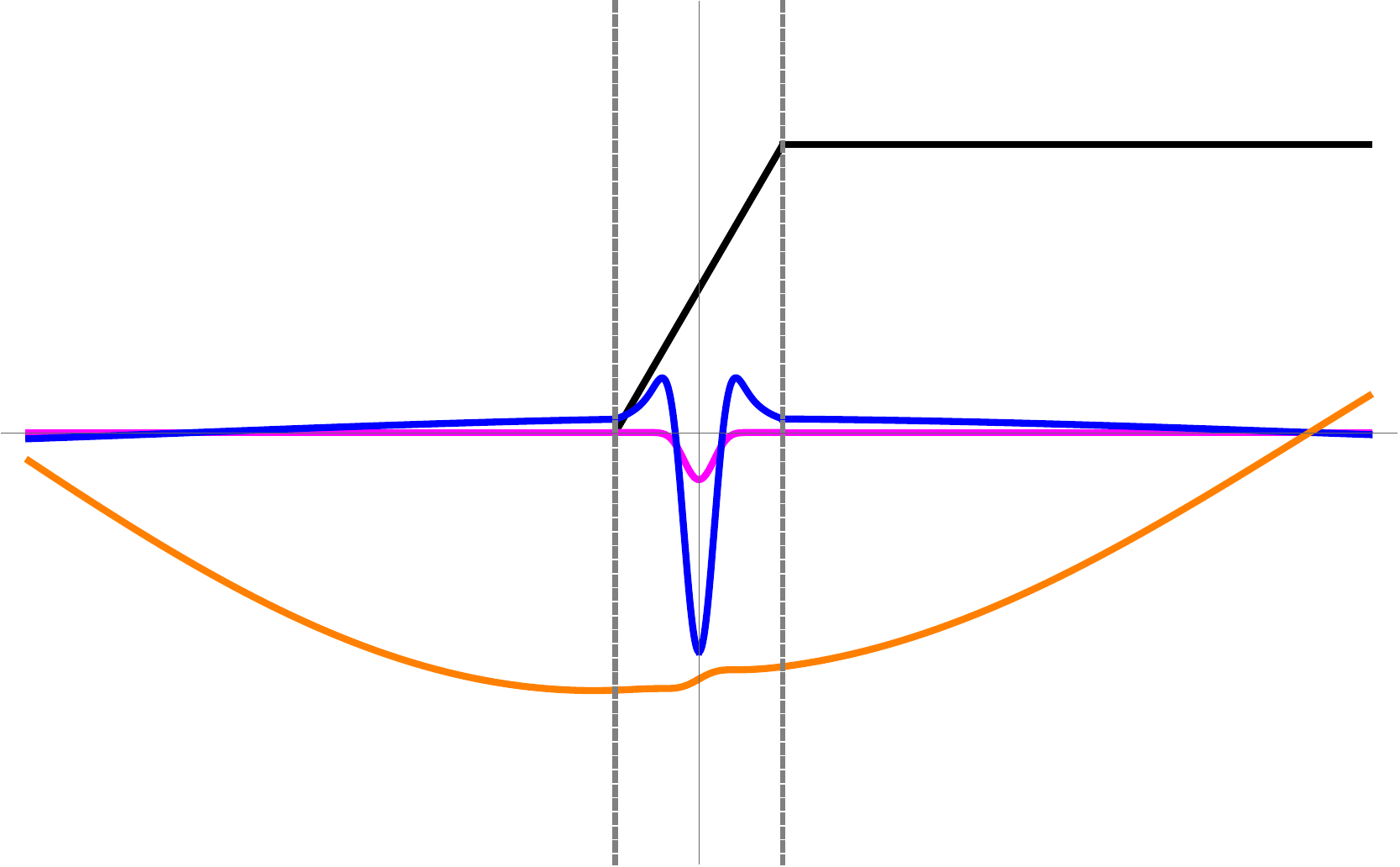}
		\caption{\small Gauge field components $a_0$ (magenta), $a_1$ (blue) and $a_P$ (orange) after removing the singular gauge transformation for the same parameters as in figure \ref{pot2}. To display all components in the same figure, $a_0$ (magenta) has been amplified and $a_1$ (blue) reduced by a factor 10.}
		\label{prepauli}
	\end{minipage}\hspace{10mm}
	\begin{minipage}{.45\linewidth}
		\includegraphics[width=6.5cm]{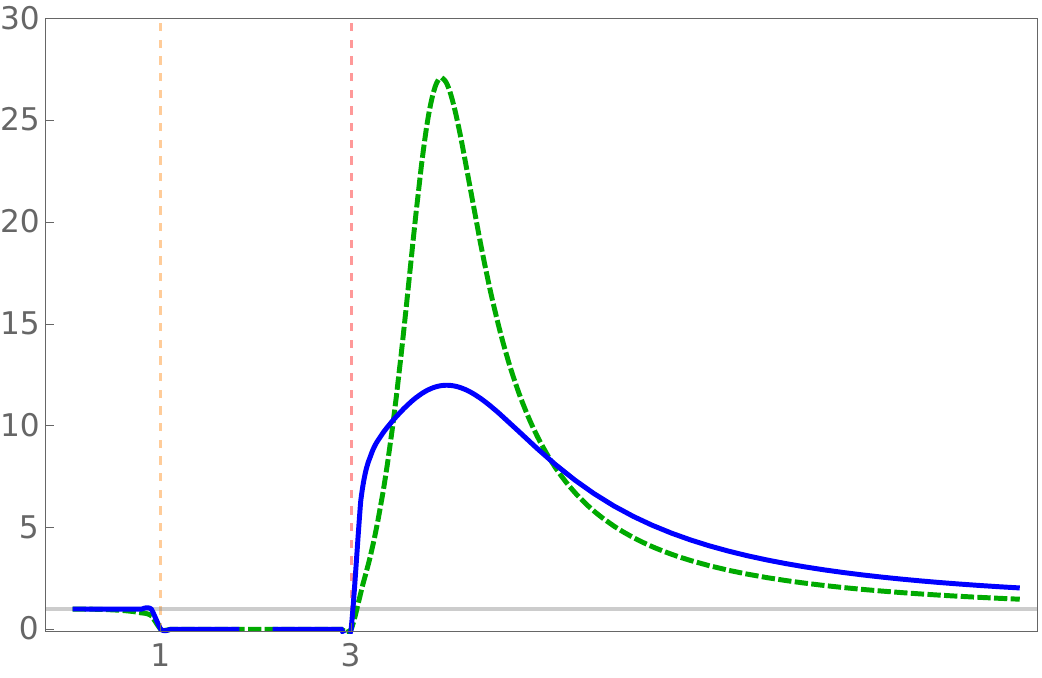}
		\caption{\small $T$ vs.\ $|E|$ for a wave packet of TM modes (blue). At the left of the orange dashed line, the propagation zone; at the right of the red dashed line, the Klein zone. We also display $T$ for TE modes (dashed green) as in figure \ref{KleinKleinTE}. We have chosen $E<0$, $k_2=0.25$, $k_\perp=0.5$, $k_0=1$ in length units of $\sqrt\theta$.}
		\label{KleinKlein}
	\end{minipage}
\end{figure}

To be more concrete, let us turn back to our background \eqref{a0}, which behaves as $\delta\!A_0=-E(x^1+L)=-Ez+k_0$ for $x^1\in[-L,L]$. Therefore, $k_0-\delta\!A_0\sim z$. For any $k_0$ in the Klein zone the coordinate $z$ vanishes at $x^1=-L-k_0/E\in[-L,L]$. As a consequence, the field $\varphi(x^1)$ scatters against a singular background of the type $z^{-2}$. Close to a singularity of this type the solutions show two different behaviors: $z^{-1}$ and $z^2$ (as shown by \eqref{sol1} and \eqref{sol2}, respectively). The regular solution \eqref{sol2} leads to finite gauge field components $a_1(x^1),a_P(x^1)$ but How do we interpret the singular gauge field that arises from the solution \eqref{sol1}? The answer is that the singular behavior comes from a singular gauge transformation. Consider a (non-singular) solution of the set of equations \eqref{mu0}-\eqref{mu2} before any specific gauge choice. In general $a_0(x^1)\neq 0$ but we can impose the temporal gauge by choosing $\alpha(x^1)$ such that $D_0\alpha(x^1)=-i(k_0-\delta\!A_0(x^1))\alpha(x^1)=-a_0(x^1)$. Therefore, in the Klein zone $\alpha(x^1)$ has a singularity at $x^1=-L-k_0/E$. As a consequence, in this gauge the time component vanishes but the spatial components $a_1,a_P$ become singular. In fact, since $\alpha$ has a simple pole at $z=0$, the transformation $a_\mu\to a_\mu+ D_\mu\alpha$ introduces a simple pole in $a_P$ and  double pole in $a_1$. This is precisely what one obtains if one inserts the solution $\varphi\sim z^{-1}$ into \eqref{a2ya1} and \eqref{a1yphi}. Fortunately, it is not necessary to abandon this gauge choice: one can undo the singular gauge transformation from \eqref{fi} to obtain a regular gauge field (figure \ref{prepauli}) or, alternatively, one can keep the singular solution but choosing the appropriate behavior at $z=0$. We choose the latter approach.

In general, one-dimensional scattering against a singularity detaches the solutions at both sides of the singular point unless a certain matching condition is specified. In other words, one could choose a certain value for the coefficients $A,B$ at one side of the singularity and a different value at the other side. If one is interested---to give an example---in solutions of definite parity, then one should choose them such that only one of the solutions \eqref{sol1} or \eqref{sol2} remains. One could even choose $A=B=0$ at one side of the singularity if it fits the physical setting. In our case, we must keep in mind that after removing the simple pole from the solution $\varphi$ one must get regular fields $a_0,a_1,a_P$. Therefore, the values of both coefficients $A,B$ must not change after crossing $z=0$. This solves the ambiguity that arises due to the singularity and permits to match the solutions \eqref{planeL} and \eqref{planeR} to determine the relation between the constants $a,b,c,d$. For this reason, the result expressed in \eqref{rl} and \eqref{tl} still holds in the Klein zone. 

Note that, as for the TE modes, $R<1$ even in the Klein zone. Of course, for a monochromatic wave with positive group velocity at $x^1>L$, one takes $c=0$ and $d=1$ in \eqref{planeL}-\eqref{planeR}, the reflection coefficient results $1/|r_L|^2$ and is thus greater than 1.
As for TE modes, this is interpreted as the result of photon production: the wave is totally reflected but a flux of photons created at the step runs in both directions.

Finally, one can also evaluate for TM modes the Klein paradox in the sense of a non-vanishing transmitted wave even for an infinitely high step. As for the TE modes, this does not occur for TM modes either. Using the standard asymptotic behavior of confluent hypergeometric functions \cite{A-S} one obtains the transmission coefficient for a wave packet for large values of the background electric field\footnote{For large $|E|$ the transmission coefficient oscillates around this expression. For simplicity, we omit the oscillatory behavior.},
\begin{align}\label{tminfinito}
	T=\frac{|t_L|^2}{|r_L|^2}\sim
	\frac{2\pi\,k_0^2}{[\Gamma(\frac34)]^2\,k^1\,\sqrt{|E|}}\,.
\end{align}
Figure \ref{KleinKlein} shows $T$ as a function of $E$.

\section{Edge states}\label{es}

For very intense electric fields such that $\theta |E|<|k_\perp|/|k^2|$ a Klein zone is created between the left- and the right-mass gaps. Propagation with energies within this zone has been discussed in the previous sections.

On the contrary, for $\theta |E|<|k_\perp|/|k^2|$ there is an overlapping between both mass gaps. If the energy lies within this overlapping then both $k^1$ and $\kappa^1$ are purely imaginary, so it could be possible to find waves localized in the vicinity of the interface $x^1=0$ (edge states). Of course, for this to happen the solution must be appropriately matched all along the $x^1$ axis.

Propagation of TE modes is described by the Klein-Gordon equation with a step-like electrostatic potential that generates an homogeneous electric field in a region of width $2L$, so states localized in the $x^1$-direction are not to be expected.

As regards the TM modes, inspection of \eqref{kg} shows that, apart from the same step-like potential, the field is subject to extra terms which depend on the incident energy $k_0$. One of these terms includes a delta-function with a negative coefficient so the equation could admit, a priori, bounded solutions.

\begin{figure}[t]
	\centering
	\includegraphics[width=13cm]{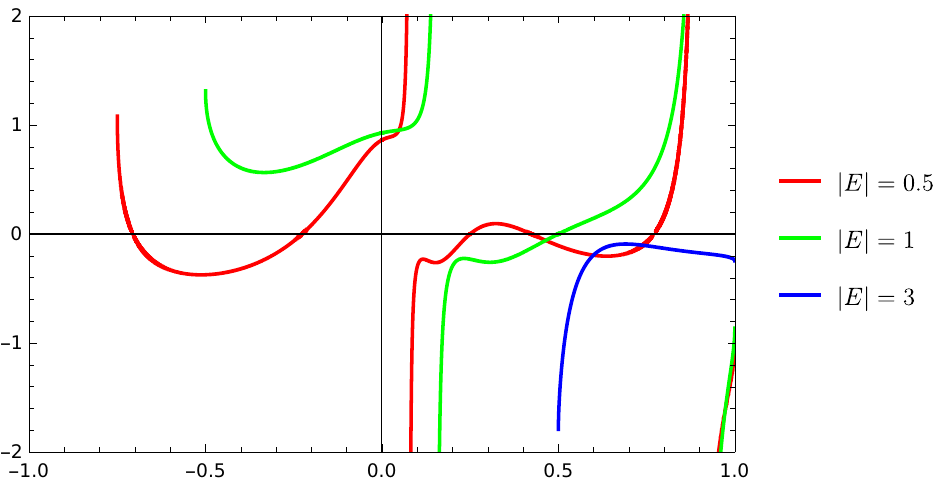}
	\caption{\small The figure shows ${\rm det}\,S$ as a function of $k_0$ for different values of the electric field ($E<0$). Each curve is defined in the interval $-2EL-k_P<k_0<k_P$, which corresponds to the overlapping of the mass gaps. The transverse momenta were chosen as $k_2=0.25$, $k_P=1$ (in units of $\sqrt\theta$.) Each intersection with the horizontal axis gives the energy of an edge state. The number of edge states decreases with increasing $|E|$.}
	\label{edge}
\end{figure}

To obtain bound states in the $x^1$-direction we look for solutions \eqref{planeL} and \eqref{planeR} with $k^1$ and $\kappa^1$ in the negative imaginary semi-axis,
\begin{align}
	k^1&=-i\sqrt{k_\perp^2-k_0^2}\,,\\[2mm]
	\kappa^1&=-i\sqrt{k_\perp^2-(k_0+2EL)^2}\,.
\end{align}
such that $b=c=0$ to match the appropriate behavior at $x^1\to\pm\infty$. This condition can only be implemented if the S-matrix is singular. In conclusion, the equation ${\rm det}\,S=0$ determines the energies at which edge states could occur. Figure \ref{edge} shows that for some values of the parameters $E,\theta,k^2,k_\perp$ there is a finite number of edge states.

\section{Conclusions}\label{conclu}

Electromagnetism in Moyal space is a self-interacting theory, so perturbations of a classical configuration do not propagate freely but interact with the underlying background. Since this interaction does not show up for homogeneous backgrounds, we have considered two half-spaces separated by a flat interface: one half is free from background fields, the other is filled up with an homogeneous electric field. One each side a photon beam propagates free from interactions but suffers scattering at the interface. In this article we have determined the stationary states, both for TE and TM modes.

Each gauge field component which is perpendicular to the plane of incidence (TE mode) decouples and propagates as a massive, charged  scalar particle. The mass is given by the momentum component parallel to the interface. Since the gauge field belongs to the adjoint representation of the $\star$-product, each TE mode only interacts with a homogeneous electric field restricted to a narrow region around the interface. As a consequence, TE modes reproduce the dynamics of a relativistic scalar field interacting with a step-like electrostatic field: $R<1$ for a monochromatic wave (but $R>1$ for a wave packet in the Klein regime) and, due to the finite width of the potential, super-Klein tunneling is suppressed and the transmission coefficients vanishes as the step height tends to infinity.

To solve for TM modes one needs to choose a particular gauge. We take the temporal gauge and solve the equations of motion. The reflection coefficient is computed and compared to the result for TE modes (figure \ref{RyT}). Also in this case, $R<1$ for a monochromatic wave (as for scalar fields; unlike fermionic fields). In the Klein regime the wave interacts with a singularity, which is an artifact of the gauge choice. A careful analysis allows one to determine the appropriate matching conditions across the singularity. After removing the singularity with a gauge transformation, we determine the profiles of the propagating gauge fields.

As the electric field goes to infinity, the transmission coefficient also vanishes. The asymptotic behavior of $T$ for TE and TM modes obeys the same power law but with a different coefficient (\eqref{teinfinito} and \eqref{tminfinito}).

For moderate electric fields, such that there is no separation between the left- and right-mass gaps, it might be possible that the electric field supports waves localized in the vicinity of the interface. Of course, this could only be the case for TM modes, for which the one-dimensional equation \eqref{kg} contains a delta-function with a negative coefficient. Figure \ref{edge} shows that for some values of the electric field a finite number of edge states may appear.

It is important to point out that the conserved current computed in this article---both for TE and TM modes---can be obtained from the asymptotic behavior of the canonical energy-momentum tensor $T_{\mu\nu}=-\frac12\{F_{\mu\rho},\partial_\nu A^\rho\}_\star+\frac14\eta_{\mu\nu}F_\star^2$ (see \cite{Grimstrup:2002xs}). As in ordinary Yang-Mills theories the construction of symmetric, traceless, gauge covariant or (covariantly) conserved energy-momentum tensors in NC gauge theories has been thoroughly studied \cite{Grimstrup:2002xs,Das:2002jd,Balasin:2015hna}. Nevertheless, the conserved current considered in the present article is only used to illustrate the asymptotic behavior of the gauge fields profiles; our main purpose is to present the classical solutions which can be subsequently used to expand the quantized gauge field.

This article therefore provides the spectrum of fluctuations around a specific background. In particular, fluctuation modes with energies in the Klein zone would indicate that inhomogeneities in the electric field are capable of generating photon beams. We think that an expansion of the quantum field in the stationary modes described in this article would provide the rate of photon creation. In the S-matrix formalism, this is usually connected with the integral of the transmission coefficient---as a function of the incident energy---in the Klein region. On the other hand, the spectral decomposition of the operator of quantum fluctuations can also be used to compute the heat-trace, from which the imaginary part of the effective action (and, thus, the stability of the vacuum) can be studied. It would be interesting to compare the results with the already studied heat-kernel methods for NC theories \cite{Vassilevich:2003yz,Bonezzi:2012vr,Ahmadiniaz:2015qaa}. Work along this two lines of research are currently under consideration.

\vspace{1cm}

\noindent{\bf Acknowledgments:} We thank Sergio Santa María Linárez for his participation in the early stages of this project, and Nicolás Grandi for useful discussions throughout this project. We also thank Fiorella Sol Beck and Maximiliano Ferro for reading the manuscript. The authors thank support from UNLP (through projects X909 and X931) and CONICET.

\vspace{.5cm}

%%%%%%%%%%%%%%%%%%%%%%%%%%%%%%%%%%%%%%%%%%%%%%%%%%

%%%%%%%%%%%%%%%%%%%%%%%%%%%%%%%%%%%%%%%%%%%%%%%%%%%%%%%%%%%%

\end{document}